\documentclass[12pt]{article}

\usepackage{graphics,epsfig}
\usepackage{amsbsy}
\usepackage{amsfonts}
\usepackage{amsmath}
\usepackage{graphicx}

\def\id{\protect{{1 \kern-.28em {\rm l}}}}

\def\k{\kappa}

\def\p{{\partial}}
\def\nn{\nonumber}

\def\dalemb#1#2{{\vbox{\hrule height .#2pt
        \hbox{\vrule width.#2pt height#1pt \kern#1pt
                \vrule width.#2pt}
        \hrule height.#2pt}}}

\let\a=\alpha \let\b=\beta \let\g=\gamma \let\d=\delta \let\e=\epsilon
\let\z=\zeta  \let\th=\theta  \let\k=\kappa
\let\l=\lambda \let\m=\mu  \let\x=\xi \let\p=\pi 
\let\s=\sigma \let\t=\tau   \let\c=\chi 
 \let\vep=\varepsilon
\let\w=\omega      \let\G=\Gamma \let\D=\Delta \let\Th=\Theta \let\L=\Lambda
 \let\P=\Pi \let\S=\Sigma  
\let\C=\Chi \let\W=\Omega
\let\la=\label \let\ci=\cite 
  
\def\nn{\nonumber} \def\bd{\begin{document}} \def\ed{\end{document}}
\def\ds{\documentstyle} \let\fr=\frac \let\bl=\bigl \let\br=\bigr
\let\Br=\Bigr \let\Bl=\Bigl
\let\bm=\bibitem
\let\na=\nabla
\def\tU{{\widetilde U}}
\let\pa=\partial \let\ov=\overline
\def\ie{{\it i.e.\ }}
\newcommand{\be}{\begin{equation}}
\newcommand{\ee}{\end{equation}}
\def\ba{\begin{array}}
\def\ea{\end{array}}
\def\ft#1#2{{\textstyle{{\scriptstyle #1}\over {\scriptstyle #2}}}}
\def\fft#1#2{{#1 \over #2}}
\def\F#1#2{{ F_{#1}^{(#2)} }}
\def\cF#1#2{{ {\cal F}_{#1}^{(#2)} }}

\def\={\, =\, }
\def\+{\, +\, }
\def\-{\, -\, }

\def\R{{\bf R}}
\def\sst#1{{\scriptscriptstyle #1}}
\def\oneone{\rlap 1\mkern4mu{\rm l}}
\def\e7{E_{7(+7)}}
\def\td{\tilde}
\def\wtd{\widetilde}
\def\im{{\rm i}}
\newcommand{\ho}[1]{$\, ^{#1}$}
\newcommand{\hoch}[1]{$\, ^{#1}$}
\newcommand{\bea}{\begin{eqnarray}}
\newcommand{\eea}{\end{eqnarray}}
\newcommand{\ra}{\rightarrow}
\newcommand{\lra}{\longrightarrow}
\newcommand{\Lra}{\Leftrightarrow}
\newcommand{\ap}{\alpha^\prime}
\newcommand{\bp}{\tilde \beta^\prime}
\newcommand{\cB}{{\cal B}}
\newcommand{\cO}{{\cal O}}
\newcommand{\vecx}{\vec{x}}
\newcommand{\vecy}{\vec{y}}
\newcommand{\vecp}{\vec{p}}
\newcommand{\vecq}{\vec{q}}
\newcommand{\tr}{{\rm tr} }
\newcommand{\Tr}{{\rm Tr} }

\newcommand{\cL}{{\cal L}}
\newcommand{\cA}{{\cal A}}
\newcommand{\cD}{{\cal D}}
\def\sst#1{{\scriptscriptstyle #1}}

\def\ve{\varepsilon}
\def\vf{\varphi}
\def\F{\Phi}
\def\wg{\wedge}

\newcommand{\wt}{\widetilde}
\newcommand{\oh}[1]{{\cal O}( #1 )}
\newcommand{\largeoh}[1]{{\cal O}\left( #1 \right)}

\def \foot {\footnote}
\def \bi{\bibitem}

\def \tr {{\rm tr}}
\def \ha {{1 \over 2}}
\def \td {\tilde}
\def \ci{\cite}
\def \N {{\mathcal N}}
\def \ww {\Omega}
\def \const {{\rm const}}
\def \ss {\sum_{i=1}^3 }
\def \t {\tau}
\def\S{{\mathcal S} }
\def \nn {\nu}
\def \XX {{\rm X}}

\def \lra {\leftrightarrow}
\def \vom {{\bar \omega}}
\def \E {{\mathcal  E}} \def \J {{\mathcal  J}}
\def \YY {{\rm Y}}

\def \d {\del}
\def \rJ {{J}}
\def \sms {sigma models\ }
\def \sm {sigma model\ }
\def \L {\Lambda}
\def \gl {\ell}
\def \tr {{\rm tr\ }}
\def\z{\zeta}
\def\zi{\zeta_1}
\def\zii{\zeta_2}
\def\K{\mbox{K}}
\def\eE{\mbox{E}}   \def \vt {\vartheta}
\def \vr {\varrho}
\def \wup {w}

\def\dg{\dagger}
\def\a{\alpha}
\def\b{\beta}
\def\e{\varepsilon}
\def\p{\phi}
\def\ap{\alpha^\prime}
\def\I{{\cal I}}

\def\R{{\bf R}}
\def\Z{{\bf Z}}
\def\C{{\bf C}}
\def\P{{\bf P}}
\def\xb{{\bar X}}
\def\Tr{{\rm  Tr}}
\def\tr{{\rm  tr}}

\def \del{\partial}
\def \a {\alpha}
\def \aa {{\a'}}
\def\g{\gamma}
\def\s{\sigma}
\def\z{\zeta}
\def\zi{\zeta_1}
\def\zii{\zeta_2}
\def\ov{\over}

\def\I{{\cal I}}
\def\J{{\mathcal J}}
\def \ok {{1\ov \k}}
\def\LL{{\mathcal L }}
\def \jL {{J}}
\def \om {\omega}
\def \cL {{\mathcal L}} \def \cH {{\mathcal H}}
\def\E{{\mathcal E}}
\def\w{\omega}
\def\b{\beta}
\def\l{\lambda}
\def\eps{\epsilon}
\def\vep{\varepsilon}
\def \De {{\mathcal D}}

 \def \cV {{\cal V}}
\def  \Jt {  {J}_{\rm tot}    }

\def \k {\kappa}
\def\foot{\footnote}
\def \four{{\textstyle {1\ov 4}}}
 \def \third { \textstyle {1\ov 3
}}
\def\det{\hbox{det}}
\def \ci {\cite}

\def \foot {\footnote}
\def \bi{\bibitem}

\def \tr {{\rm tr}}
\def \ha {{1 \over 2}}
\def \tid {\tilde}
\def \vv {{\rm v}}
\def \tl {{\tilde \l}}
\def \XX {{\rm X}}
\def \ta {{\tilde \a}}
\def \fo { {1\ov 4}}
\def \ep {\epsilon}
\def \inti {{\int^{2\pi}_0 {d \sigma \ov 2 \pi}}}

\def \d {\partial}
\def \K {{\rm S}}
\def \el {\ell}
\def \Tr {{\rm Tr}}
\def \P {\Phi}
\def \l  {\lambda}
\def \tl {{\tilde \l}}
\def \bl {{\tilde \l}}
\def \const {{\rm const}}
\def \V {v}

\def \bv {v^*}
\def \vv {{\rm v}}
\def \LL {{\mathcal L}}
\newcommand{\PV}[1]{P_{\!\!_{V_{#1}}}}

\def \bL {\ell}
\def \M {{\mathcal M}}
\def \N {{\mathcal N}}
\def \S {{\rm S}}
\def \vn {\vec n}
\def \tl {\td \l}
\def \td {\tilde}
\def \Prod {\Pi}
\def \O {{\mathcal O}}
\def \Q {{\rm  Q}}
\def \D {\Delta}
\def \N {{\mathcal N}}
\def\tN{{\tilde N}}

\def \m {\mu}
\def \vs {\vec \s}
\def \ie {i.e.}

\def \cD {{\cal D}}

\def  \le  {\l_{\rm eff}}

\def \rS {{\rm S}}
\def\as{{\a}}
\newcommand{\bra}[1]{\mbox{$\langle #1 |$}}
\newcommand{\ket}[1]{\mbox{$| #1 \rangle$}}

\newcommand{\auth}{AUTHORS}

\def\thb{\bar{\theta}}
\def\Thb{\bar{\Theta}}
\def\barp{\bar{p}}
\def\barq{\bar{q}}
\def\barc{\bar{c}}
\def\bard{\bar{d}}
\def\e{\epsilon}

\def \bi{\bibitem}
\def \la {\label}

\def \l {\lambda}
\def\foot{\footnote}
\def \tl  {{\tilde \l}}
\def \sql {{\sqrt \l}}
\def \adss {$AdS_5 \times S^5$\ }
\newcommand{\rf}[1]{(\ref{#1})}
\def \ov {\over}

\def\th{\theta}
\def\Th{\Theta}
\def\vth{\vartheta}

\def\vth{\vartheta}

\def\ra{\rightarrow}
\def\N{{\cal N}}
\def\F{{\cal F}}
\def\cc{\circ}
\def\eqv{\equiv}

\def\ni{\noindent}
\def \ha{{1\ov 2}}
\def \bw {{\rm w}}

\def\r{{\rm r}}

\def \cT {{\cal T}}
\def \no {\nonumber}

\def \J {\mathcal{J}}
\def \del {\partial}

\def \bps {{\bar \psi}}
\def \sqbl {\sqrt{\bar \lambda}}

\def\dF{\dot{F}}
\def\dG{\dot{G}}
\def\df{\dot{f}}
\def \E {{\cal E}}

\def \S {{\cal S}}
\def \J {{\cal J}}

\def\ms{\mathcal{S}}
\def\mj{\mathcal{J}}
\def\soj{\fr{\ms}{\mj}}
\def \R {{\bf R}}
\def \om {\omega}
\def \tH {\widetilde H}
\def \bE {\bar E}
\def \x {{\cal X}}

\def \hV {{\hat V}}

 \def \bb {\bar \beta}
\def \W {{\cal E}}

\def \bi{\bibitem}
\def \la {\label}

\def \l {\lambda}
\def\foot{\footnote}
\def \tl  {{\tilde \l}}
\def \sql {{\sqrt \l}}
\def \sqtl {{\sqrt {\tilde \l}}}

\def \HH {{\rm E}}
\def \cS {{\cal S}}
\def \cL {{\cal L}}

\def \adss {$AdS_5 \times S^5$\ }

\def \D {\Delta}
\def \thet {\theta}
 \def \t {\tau}
 \def \p {\phi}
 \def \r {\rho}
 \def \rN {{\rm N}}
 \def\tw{{\tilde w}}
 \def\hJ{{J}}
 \def\hw{{w}}
 \def\hl{{\lambda}}
 \def\hth{{\theta}}
 \def\NN{{\cal N}}
 \def \bv {{ \bar w}}
\def \vn {{\vec n}}
\newcommand{\sfrac}[2]{{\textstyle\frac{#1}{#2}}}
\def \bl {{ \bar \lambda}}
\def \bp {{\bar p}}
\def \bu {{\bar u}}
\def \sha {\sfrac{1}{2}}
\def \w {\omega}
\def \ov {\over}
\def \vl { \vec \ell}
\def \varpi {{\rm w}}
\def \OO {{\cal O}}
\def \bG {\bar \G}

\def \c {\gamma}

\def \ss {{\rm s}}

\def \ve {\varepsilon}
\def \pa{\partial}
\def \I {{\cal I}}
\def \LL {{\cal L}}
\def \ep {\epsilon}
\def \R {{\rm R}}
\def \tilt {{\tilde t}}

\def\pic #1#2{\hbox{\lower#1pt\hbox{~\mbox{\epsfxsize=20truemm \epsffile{#2}}}}}

\def\pic #1#2#3{\hbox{\lower#1pt\hbox{~\mbox{\includegraphics[scale=#3]{#2}}}}}

\def \bt {\bar\theta}
\def \te {\theta}

\def \cc {{\rm f}}
\def \d {\delta}

\def \cL {{\cal L}}
\def \S  {{\cal S}}
\def \pp {{q}}
\def \vt {\vartheta}
\def \mm {{\cal  \ell}}
\def \Z {{\cal Z}}
\def \pa {\partial}
\def \C {{\cal C}}
\def \be {\bea}
\def \ee {\eea}
\def \c {\gamma}  \def \d {\delta}
\def \eps {\epsilon}

\def \bp {\begin{pmatrix}}  \def \ep {\end{pmatrix}}

 \def \T {{\cal T}}

\def \bp {\begin{pmatrix}}  \def \epm {\end{pmatrix}}
\def \ha {{\textstyle{1 \ov 2}}}

\begin{document}
\overfullrule=0pt
\parskip=2pt
\parindent=12pt
\headheight=0in \headsep=0in \topmargin=0in \oddsidemargin=0in

\vspace{ -3cm} \thispagestyle{empty} \vspace{-1cm}
\begin{center}
 \vspace{2cm}
{\Large\bf
 On the short string limit of        \\
\vspace{0.3cm}
the folded spinning string in $AdS_5 \times  S^5$
 }

 \vspace{.8cm} {
  M. Beccaria$^{a,}$\footnote{matteo.beccaria@le.infn.it} and
 A. Tirziu$^{b,}$\footnote{atirziu@purdue.edu} }\\
 \vskip 0.1cm

{\em
$^{a}$
Physics Department, Salento University and INFN, 73100 Lecce, Italy \\
\vskip 0.04cm
$^{b}$ Department of Physics, Purdue  University,
W. Lafayette, IN 47907-2036, USA\\
 }

\end{center}

 \vskip 0.8cm

 \begin{abstract}

In this paper we generalize the results of
arXiv:0806.4758   to non-zero  value $J$ of angular momentum in $S^5$. We compute the $1$-loop
correction to the energy of the folded spinning string in $AdS_5 \times S^5$ in
the particular limit of slow short string approximation. In this limit the string is moving
in a near-flat central region of $AdS_5$ slowly rotating in both $AdS_5$ and $S^5$. The
one-loop correction should represent the first subleading correction to strong coupling
expansion of the anomalous dimension of short gauge theory operators of the form $\mbox{Tr}\, D^S \Phi^J$
in the $SL(2)$ sector.

\end{abstract}
\newpage


\renewcommand{\theequation}{1.\arabic{equation}}
 \setcounter{equation}{0}

\setcounter{equation}{0} \setcounter{footnote}{0}
\setcounter{section}{0}

\section{Introduction}

The folded rotating string moving in
$AdS_3$ \ci{gkp,dev} was extensively used recently to uncover the integrable structure
underlying the spectrum of planar $\N=4$  SYM theory and the free \adss superstring theory.
The classical  energy of this configuration is proportional to the string tension, i.e.
$E_0 = \sql\ \E( \S)$, where $\S$ is the scaled spin $\S= { S \ov \sql} $. The particular limit that proved to be very useful in connection
to gauge theory, is the large $\S$ limit where one finds at leading order \ci{gkp}:
$E_0 = S + { \sql \ov \pi} \ln S + \mbox{subleading terms}$. The extension beyond the $\ln S$ order was carried out in 
\cite{Basso:2006nk,bftt}, and
precisely the same structure was obtained on both string and gauge theory side in the large $S$ limit.

For the folded spinning string solution, the  radial coordinate $\r$ of the global $AdS_5$ space
($ds^2 = - \cosh^2 \r\ dt^2 + d\r^2 + \sinh^2 \r \ d\Omega_3^2$)  is
expressed in terms of  an elliptic function of the spatial string coordinate
$\s$  and thus finding the explicit   form of  the 1-loop
correction \ci{ft1} to the energy $E_1$
of this soliton solution of 2d  string sigma model
appears to be  technically challenging.
The analytic form of the quantum correction can be found
 in the limit of large $\S$  when
the ends of the string
reach the boundary of the $AdS_5$. Then the  solution drastically
simplifies ($\rho$ becomes piece-wise
linear in $\s$)  \ci{ft1,ftt}  and   one finds   that
 $E_1= c_1 \ln S + ..., \ \  c_1= - 3 \ln 2$. Beyond the $\ln S$ order, the ``long string'' no longer touches the boundary
 but it is close to it, and one can still compute the $1$-loop correction to energy \cite{bftt}.

The folded spinning string solution with non-zero spin $J$ in $S^5$, and large $S$ limit was also studied.  The comparison of
the strong-coupling string theory expansion in the limit
$\l \gg 1$  with $   \J = { J \ov \sql}, \ \S= {S \ov \sql }$, \
$ \ell \equiv { \J\ov  \ln \S}$=fixed and $\ell   < 1$ with the weak coupling gauge theory expansion
of anomalous dimensions of sl(2) sector operators
in the limit $\l \ll 1$ with $  \ J \gg 1, \ S \gg 1$, \
$j= { J \ov \ln S}$=fixed and $ j < 1$   was analyzed in detail in  \ci{ft1,ftt, bfst,bgk,am,rt,frs}.

Most of the remarkable progress achieved in understanding the spectra of free \adss superstring theory and
of planar $\N=4$  SYM theory was restricted to a sector of gauge theory operators with large number of
fields/derivatives or strings with large values of  quantum numbers like spins. It is  important  to try
to learn more about dimensions/ energies of ``short''operators/strings. For this purpose it would be very
useful to understand the Bethe Ansatz beyond the asymptotic regime. Another step is to study the energy of strings
carrying parametrically small values of spins.

Motivated by the desire to shed light on this issue, in \cite{tt} it was
studied the folded spinning string in the small $\S$ limit. Since rotation of the string balances the contracting
effect of its tension, smaller values of the spin correspond to smaller values of the
length of the string whose  center of mass is at $\r=0$. $\S$ essentially measures the
length of the string. Having in view the success of ``long folded'' string in describing the corresponding minimal gauge theory
operator $\mbox{Tr}\,( D^S \Phi^J)$ in the large $S$ limit, one is tempted to conjecture that the ``short string'' limit describes
the operator with finite $S$.

Since the $AdS_5$  space is nearly flat at  the vicinity
of $\r=0$, the slowly rotating   (i.e. small) string
with $\S \ll 1$    should  have essentially the same classical
 energy as in flat space \ci{gkp}, i.e.
$E_0 = \sqrt{ 2 \sql S} + O(S^{\frac{3}{2}})$. This is indeed the case as was shown in \cite{tt}.
The $1$-loop correction of the first two leading terms in the  small spin  expansion
were found explicitly in \cite{tt}. In this paper we consider the generalization of the ``slow string" limit by adding a
small spin $ \J= \frac{J}{\sqrt{\lambda}}$ in $S^5$. More precisely, we study both the classical energy and $1$-loop quantum corrections to the energy
of the slowly rotating ``short string" $\S \ll 1$, $\J \ll 1$ with $u\equiv \frac{\J^2}{\S}= \frac{J^2}{S \sqrt{\lambda}}$ {\it fixed}. This generalization is potentially important as it allows one to relate the corresponding
string states to operators like $\Tr (D_+^S \Phi^J)$
in the closed $sl(2)$  sector of the SYM theory  (with $J$ having the interpretation
of the  length of the corresponding spin chain \ci{bs}).

Our results can be summarized as follows. Given  the energy  $E(S,J, \l) $ of the corresponding
state in the AdS/CFT spectrum we may  expand it at large $\l$  with $\S = { S \ov \sql}$ and $\J = { J \ov \sql}$
fixed, i.e. in the semiclassical string limit. Expanding {\it then} in  the limit $\S \ll 1$, i.e.
$S \ll \sql $, with $u= \frac{J^2}{S \sqrt{\lambda}}$ fixed\footnote{This scaling is implied by the classical energy $\mathcal{E}_0=\sqrt{\J^2+2 \S \sqrt{1+\J^2}}+...$. We consider the limits with both $\S$ and $\J$ small.} and reexpressing $E$ as a function of $S$, $J$ and $\l$  one is to find
\bea
&& E(S,J,\lambda)=
 \lambda^{{1}/{4}}\sqrt{S}\ \sqrt{u+2}\Big[h_0(\lambda,u) + h_1(\lambda,u)S + h_2(\lambda,u) S^2 + ...\Big]\ ,
\    \label{lmt} \\
&&  h_n(\lambda,u) =  \frac{1}{(\sqrt{\lambda})^n}[ a_{n0}(u) + \frac{a_{n1}(u) }
{\sqrt{\lambda}}+\frac{a_{n2}(u)}{(\sql)^2}+...] \ .
 \la{ha}
\eea
In the classical   string theory limit
\be
a_{00}(u)=1\  , \ \ \ \ \  \ \ a_{10}(u)= \frac{2 u+3}{4 (u+2)}  \ , \ \ \ \ \ \ \ a_{20}=-\frac{2 u[2u (u+5)+17]+21}{32 (u+2)^2}  \ , ... \ee
while our 1-loop string computation gives
\begin{eqnarray} \la{on}
a_{01}(u)=0\  , \ \ \ \ \ \ \ \  a_{11}(u)=\frac{41-8 u (u-1)-32 \zeta(3)}{32 (u+2)} , \nonumber \\
\ a_{21}(u)=\frac{A_0+A_1 u+A_2 u^2+A_3 u^3 +A_4 u^4}{32 (u+2)^2}  \ , ...
\end{eqnarray}
where
\begin{eqnarray}
A_0&=& -7566 + 5344 \zeta(3)+3840 \zeta(5), \quad A_1=-10671+7504 \zeta(3)+4992 \zeta(5),\nonumber\\
A_2&=&-4425 + 3408 \zeta(3)+576 \zeta(5), \quad
A_3= -96 +736 \zeta(3)- 672 \zeta(5), \nonumber\\
 A_4&=& 192 +96 \zeta(3)-96 \zeta(5) \ .
\end{eqnarray}

Since the leading $\lambda^{{1}/{4}}\sqrt{S}\ \sqrt{u+2}$   term is essentially the same as that of a folded spinning string solution in
the flat-space string  theory,
it is natural to conjecture, as in \cite{tt}, that $h_0(\lambda,u)=1$ to all orders in string coupling expansion.
This was already confirmed by $1$-loop computation in \cite{tt}. One may, of course, consider formally any particular limit of $u$ in the string expression (\ref{lmt}). In particular the small $u$ limit represents the situations when semiclassical spin in $S^5$ is much smaller then the corresponding spin in $AdS_5$, while large $u$ limit is the other way around\footnote{In both cases both semiclassical spins $\J, \S$ are small, i.e. $\J, \S \ll 1$.}.

The strong coupling string theory expansion (\ref{lmt}) could  be naively compared with the weak coupling gauge theory expansion for the anomalous dimension
of operator $ \mbox{Tr}\, D^S \Phi^J$ in the $sl(2)$ sector in the limit $\lambda \ll 1$ with integer fixed spins $J, S$, and then analytically continued to small $S$
with $l \equiv \frac{J^2}{S}$=fixed. However, as in the case with $J=0$ (see below) or the case of a similar expansion in the large $S$ limit with fixed $\frac{\J}{\ln \S}$,
one expects to need a resummation in all arguments ($\lambda, S,J$) in order to compare the string and gauge theory expansions.

To recover the $J=0$ limit in the short string case we take $u=0$ in (\ref{lmt}). Then the energy is
\bea
 E(S,\lambda)=
 \lambda^{{1}/{4}}\sqrt{2 S}\ \Big[1  +    (a_{10}   +
  \frac{a_{11}}{\sqrt{\lambda}} + ...) { S \ov \sql }     +  (a_{20}+ \frac{a_{21}}{\sqrt{\lambda}}+ ...)\frac{S^2}{\lambda}+ O(S^3) \Big]
  \label{mt} \  , \ee
where
\begin{eqnarray} \la{cl}
a_{10}&=& { 3 \ov 8}  \ , \ \ \ \ \ a_{20}=-{21\ov 128}  \ ,  \ \ \ \ \ a_{11}= { 41 \ov 64}  - \ha \zeta(3) \approx 0.039 \ ,  \nonumber\\ a_{21}&=&-\frac{1261}{1024}+\frac{167}{192}\zeta(3)+\frac{5}{8}\zeta(5) \approx 0.4622 \ .
\end{eqnarray}
The coefficient $a_{11}$ was already obtained in \cite{tt}. Here we compute the $1$-loop correction to order $S^{\frac{5}{2}}$. Interestingly, at least to this order, only $\zeta$-functions of odd argument appear. 
As already mentioned, the structure of this small $S$ expansion cannot be easily compared with a gauge theory result\footnote{This is in contrast to the large spin or ``long string''  limit   where
the limits of large $\l$ and large $S$ appear to commute. The
perturbative string theory  and perturbative gauge theory limits are
in fact different as limits of functions on
 the two-parameter space $(\l, S)$:
 in string theory  one assumes  $\l \gg 1$
with $\S= {S \ov \sql}$ fixed and then takes $\S$ large for long string, or $\S$ small for short string;
in gauge theory one assumes  $\l \ll 1$
with $S$ fixed and then takes $S$ large for long string, or keeps it finite or formally expand in small $S$ for short string. }.
In the gauge theory, the spin $S$ is an integer and its analitical continuation to small values is a very difficult problem due to the intrinsic
dependence of the mixing structure on the integer $S$ as explained 
in~\cite{Braun:1999te} (see also \cite{Korchemsky:2003rc} for a fully understood $S\to -1$ continuation). A remarkable simplification occurs in the 
special case of twist-2 operators $\tr ( \Phi   D_+^S \Phi)$ belonging to the $SL(2)$ sector, where $\Phi$ is complex scalar, computed
at fixed $S$ in weak coupling perturbation theory. In this case, the anomalous dimension of the scaling field is known as a closed function of the spin (see, e.g.,
\ci{kot})  and can be analytically continued to small $S$.
\begin{equation}
\gamma( \lambda, S) =  q_1(\lambda)  S + q_2(\lambda) S^2 +  O(S^3) \ , \ \ \ \ \ \ \ \        \l \ll 1, \ \
      \ ,   \label{ak}
\end{equation}
where
\begin{equation} \la{uy} q_1(\lambda)=  d_{11}  \lambda + d_{12}  \lambda^ 2
+ ... \ , \ \ \ \ \ \
\
q_2(\lambda)= d_{21}  \lambda + d_{22}  \lambda^ 2 + ... \ , ....
\end{equation}
However, even in this case, to relate the  ``small spin''  string theory  \rf{mt}  and gauge theory
\rf{ak} expansions one would need to resum the series in both arguments $(\l,S)$.
It would be certainly very interesting if one can resum the series on at least one side of the duality, and
then be able to effectively compare the two sides.

\
The rest of the paper is organized as follows.
We start  in section 2 with a review of the folded spinning
string solution in $AdS_5  \times S^5$ and its small spins expansion. We also consider a similar
solution with two spins in flat space.

In section 3 we shall first  recall the general expression
for the quadratic fluctuation Lagrangian $\widetilde {\cal L}$ \ci{ft1}
of the \adss superstring \ci{mt}  near the folded spinning
string solution.  We also review the method used in \cite{tt} to compute the $1$-loop correction. In subsection 3.1 we expand
the coefficients in $\widetilde {\cal L}$   in the small spin or short string
parameter $\eps$, and explicitly compute the $1$-loop correction at order $O(\epsilon^4)$. The $\epsilon$ expansion
may be viewed as a  particular case of a near flat space expansion  of the quantum \adss
superstring. As was already found in \cite{tt}, we indeed observe that the
leading $O(\eps)$  term in the 1-loop  string energy vanishes.

In subsection 3.2 we shall expand the  2d determinants  that enter the
expression  for the 1-loop partition function further at
$O(\epsilon^6)$ order in $\eps$, and explicitly compute the
$1$-loop correction to the short string at order $O(S^{\frac{5}{2}})$.

In Appendix A we present some details of the contributions from various fields to the $1$-loop correction at order $O(\epsilon^4)$.

In Appendix B we show details of the exact computation of the integrals that appear after summations. Remarkable, the integrals can be done exactly, and thus $1$-loop coefficients can be obtained exactly.

\renewcommand{\theequation}{2.\arabic{equation}}
 \setcounter{equation}{0}

\setcounter{equation}{0} \setcounter{footnote}{0}

\section{Short string limit  of   folded spinning string solution}

Let us start with a review of the classical solution for the
folded  string spinning moving in $AdS_5 \times S^5 $,
\begin{equation}\la{so}
 t= \kappa \tau, \quad \phi= w \tau, \quad \varphi= \J \tau, \quad \rho=\rho(\sigma) \ , \ \ \ \ \ \ \
ds^2= -\cosh^2 \rho\ dt^2 + d \rho^2 + \sinh^2 \rho\ d \phi^2 + d \varphi^2
\ , \end{equation}
where
\begin{equation}
\rho'^2 = \kappa^2 \cosh^2 \rho - w^2 \sinh^2 \rho \-\J^2 .   \label{snh}
\end{equation}
$\rho$ varies from $0$ to its maximal  value $\rho_*$
\begin{equation}
\coth^2 \rho_* = \frac{w^2-\J^2}{\kappa^2-\J^2}\equiv 1+ \frac{1}{\epsilon^2} \ ,
\end{equation}
where as in \cite{tt} we introduced a parameter $\epsilon$. Here $\J = \frac{J}{\sqrt{\lambda}}$ plays the role of the semiclassical $S^5$
momentum parameter and $\eps$ measures the length of the string; indeed one can see this by expanding $\r_*= \eps - { 1 \ov 6}{ \eps^3} + ... $ in small $\epsilon$. The solution of the  differential equation (\ref{snh}), i.e.
\begin{equation}
\rho' = \pm \sqrt{\kappa^2-\J^2} \sqrt{1-\eps^{-2}\, {\sinh^2 \rho}}\ , \quad \quad \rho(0)=0
\end{equation}
 can be written in terms of a  Jacobi elliptic function
\begin{equation}
\sinh \rho=\epsilon \ {\rm sn}({\sqrt{\kappa^2-\J^2} \eps^{-1}  \sigma},\ -\epsilon^2) \ . \label{mlq}
\end{equation}

The periodicity condition, and the charges are
 \bea\la{pion}
&& \sqrt{\kappa^2-\J^2}=\epsilon \
 _2 F_1\left(\frac{1}{2},\frac{1}{2};1;-\epsilon^2\right) \ , \  \ \ \  \
  \ \ \mathcal{E}_0\equiv \frac{E_0}{\sqrt{\lambda}}=\frac{\kappa}{\sqrt{\kappa^2-\J^2}}
\epsilon \
  _2F_1\left(-\frac{1}{2},\frac{1}{2};1;-\epsilon^2\right) , \nonumber \\
&&
 \mathcal{S}\equiv \frac{S}{\sqrt{\lambda}}=\frac{w}{\sqrt{\kappa^2-\J^2}}\frac{\epsilon^3}{2}
 \ _2F_1\left(\frac{1}{2},\frac{3}{2};2;-\epsilon^2\right) \ , \quad \quad J= \J \sqrt{\lambda} \ .
\eea

To consider the short string limit we expand  as in \cite{tt} in small $\eps$ while keeping
 $\J$  arbitrary.  Then  we find
 \begin{equation}
 \mathcal{E}^2_0=
\J^2 + \epsilon^2 (1+\J^2) + \frac{\epsilon^4}{2}\left(1+\frac{\J^2}{4}\right)-\frac{\epsilon^6}{32}+\left(\frac{1}{64}-\frac{5 \J^2}{1024}\right)\epsilon^8+O(\epsilon^{10})
\end{equation}
\begin{equation}
\S^2 = \frac{\epsilon^4}{4} (1 + \J^2) + \frac{\epsilon^6}{16} (1 - \J^2) + \frac{9}{256}(\J^2-1) \epsilon^8+
O(\epsilon^{10})
\end{equation}
i.e.
\begin{equation}
\epsilon^2 =\frac{2 \mathcal{S}}{\sqrt{1 +\J^2}}  + O(\S^2) \ , \ \ \ \ \ \ \ \ \ \
 \mathcal{E}^2_0 = \J^2 + 2 \mathcal{S}\sqrt{1+\J^2}   + O(\S^2)  \ . \la{maq}
 \end{equation}
The short string limit $\eps \ll 1$  \ci{ft1}
can   thus be achieved  by, e.g.,
  considering  a slowly spinning string $\S \ll 1$  or by
assuming large momentum
in $S^5$, i.e.  $\J \gg 1$.
 In this paper we concentrate in the slow short string limit where we have $\eps \ll 1, \ \S \ll 1$ and $\J \ll 1$.
Fixing the ratio $u \equiv \frac{\J^2}{\S}$ and taking the limit of small $S \ll 1 $, the parameter $\epsilon$ and the classical
energy are expressed as
\begin{eqnarray}
\epsilon &=& \sqrt{2 \S}- \frac{2 u+1}{4 \sqrt{2}}\S^{\frac{3}{2}}+\frac{27+ 4 u (5 u+11)}{64 \sqrt{2}}\S^{\frac{5}{2}}+O(\S^{\frac{7}{2}})\label{zz2} \ , \\
\mathcal{E}_0&=&\sqrt{u+2}\sqrt{\S}+\frac{(2 u+3)}{4 \sqrt{u+2}}\S^{\frac{3}{2}}-\frac{21+2 u [17+2 u (u+5)]}{32 (u+2)^{\frac{3}{2}}}\S^{\frac{5}{2}}+O(\S^{\frac{7}{2}}) \ . \label{zz1}
\end{eqnarray}
In section 3 we will find the $1$-loop correction to this classical energy for arbitrary $u$.

\bigskip

The above small spin expansion (\ref{zz1}) is an example of
 a near flat space expansion: the
 leading-order in $\S$  energy
can  be identified with the  energy of folded  spinning  string solution in the flat space
\begin{equation}
t= k  \tau\ , \quad \rho=\eps  \sin \sigma\ , \quad \phi=\tau\ , \quad \varphi = \J \tau, \ \ \ \ \ \
ds^2=-dt^2 + d \rho^2 +\rho^2 d \phi^2 + d \varphi^2 \ ,
\end{equation}
where here $\eps$ is an arbitrary  constant amplitude.
The energy and the spin  then satisfy the usual flat-space  Regge relation
(we use  string  tension $T= {\sql \ov 2 \pi}$)
\begin{equation}
E_0 = \sqrt{\epsilon^2+\J^2} \sqrt{\lambda}\ , \qquad S=   \frac{\eps^2}{2}\sqrt{\lambda} \ ,
\ \ \ \ \ {\rm i.e.} \ \ \ \ \ \
\E_0 = \sqrt{2 \S +\J^2}  \ . \label{mbj}
\end{equation}
In the flat space case  this is the  exact expression for any values of the spins $S$ and $J$,
which also does not receive quantum corrections.

\bigskip

We can further consider particular limits of small and large $u$ respectively in the energy (\ref{zz1}).  To make the connection to the $\J=0$ case let us expand the classical energy in the limit when the $S^5$ rotational energy is smaller than the spinning  one, i.e.  $ \J \ll \sqrt \S \ll 1$. In this limit $\epsilon$ is expressed as
\begin{equation}
\epsilon=
 \sqrt{2 \mathcal{S}}-
 \frac{1}{4 \sqrt{2}}\mathcal{S}^{{3}/{2}} \left(1 +\frac{2\J^2}{\S}\right)
 + \frac{27 }{64 \sqrt{2}}\S^{\frac{5}{2}}\left(1+ \frac{44}{27}\frac{\J^2}{\S}+\frac{20}{27}\frac{\J^4}{S^2}\right) + O(\S^{\frac{7}{2}}) \ ,
 \end{equation}
and the classical energy  has the following expansion
 \begin{eqnarray}
\mathcal{E}_0 &=& \sqrt{2 \mathcal{S}}\ \left(1+ \frac{\J^2}{4\mathcal{S}}
+...\right)
+\frac{3}{4 \sqrt{2}}\mathcal{S}^{{3}/{2}}\left(1+ \frac{5\J^2}{12\mathcal{S}}
+...\right) + \nonumber \\
&& - \frac{21}{64 \sqrt{2}}\S^{\frac{5}{2}}\left(1+ \frac{73}{81} \frac{\J^2}{\S}+...\right)+ O(\S^{\frac{7}{2}}).
   \la{zz}
  \eea

The other limit of large $u$ means  $\sqrt \S \ll \J \ll1$. In this limit the classical energy (\ref{zz1}) can be expanded as ($E_0 = \sqrt{\lambda}\mathcal{E}_0$)
\begin{eqnarray}
E_0&=&J \bigg(1+ \sqrt{\lambda}\frac{S}{J^2}- \lambda \frac{S^2}{2 J^4} + ...\bigg)+ \frac{S J}{2 \sqrt{\lambda}}\bigg(1+ \sqrt{\lambda}\frac{S}{2 J^2}- \lambda^{\frac{3}{2}}\frac{S^3}{4 J^6}+...\bigg) \nonumber\\
&-& \lambda^{-\frac{3}{2}} \frac{S J^3}{8}\bigg(1+ 2 \sqrt{\lambda}\frac{S}{J^2}+ \lambda \frac{S^2}{J^4}+...\bigg)+O(S J^5) \label{zz3}
\end{eqnarray}
Let us note that, despite the expectation that this is a BMN type limit, i.e. energy going in powers of $\lambda$, this is in fact not the case.
This difference appears because in order to obtain (\ref{zz3}) we take the limit of small $\S$ with $u$ fixed, and then large $u$, while in the BMN case one should take the limit of large $\J$ first with $\frac{\J}{\S}$ fixed, and then taken large.

\renewcommand{\theequation}{3.\arabic{equation}}
 \setcounter{equation}{0}

\section{$1$-loop correction to the energy}

Expanding the $AdS_5 \times S^5 $ superstring action in conformal gauge to quadratic order in the fluctuations near the folded spinning string
solution one finds  \cite{ft1}
\be
\widetilde S = -\frac{\sqrt\lambda}{4\,\pi}\int d\tau\,\int_0^{2\,\pi} d\sigma\,(\widetilde {\cal L}_B + \widetilde {\cal L}_F),
\ee
where the bosonic quadratic fluctuation Lagrangian is
\begin{eqnarray}
\widetilde {\cal L}_B &=& - \partial_a \td {t} \partial^a \td {t}- \mu_t^2 \td {t}^2 +   \partial_a \td {\phi} \partial^a \td {\phi}+ \mu_{\phi}^2 \td {\phi}^2\nonumber\\
&+& 4 \td {\rho} (\kappa \sinh \rho\ \partial_0 \td {t} - w \cosh \rho\ \partial_0 \td {\phi})+  \partial_a \td {\rho} \partial^a \td {\rho}+\mu_{\rho}^2 \td {\rho}^2\nonumber\\
&+& \partial_a {\beta}_u \partial^a {\beta}_u +\mu_{\beta}^2 {\beta}_u^2 +
 \partial_a {\varphi} \partial^a {\varphi}+\partial_a {\chi}_s \partial^a {\chi}_s + \J^2 \chi_s^2 \ ,   \label{lag}
\end{eqnarray}
where
\bea
&&\mu_t^2= 2 \rho'^2 -\kappa^2 +\J^2,
 \qquad \mu^2_{\phi}=2 \rho'^2 -w^2 +\J^2,
 \qquad \mu^2_{\rho}=2 \rho'^2 -w^2-\kappa^2+2 \J^2, \no \\
 && \mu_{\beta}^2=2 \rho'^2 +\J^2   \ , \qquad \mu_{F}^2 = \rho'^2 +\J^2 \ .
 \label{laes}
\eea
The two bosons $\beta_i$ ($i=1,2$) are two $AdS_5$ fluctuations transverse to the $AdS_3$ subspace in which the string is moving, while the
$\varphi,\chi_s$ ($s=1,2,3,4$) are fluctuations in $S^5$.

The fermionic lagrangian describes a system of 4+4 2d Majorana fermions
\be
\wt {\cal L}_F = 2\,i\,(\overline\Psi \gamma_a \partial^a\,\Psi-\mu_F\,\overline\Psi\,\Gamma_{234}\,\Psi).
\ee
Rotating to Euclidean time, $\tau \to i \tau$, and introducing the infinite time interval $\T$, we can write
\be
\wt {\cal L}_B = (\wt t, \wt \phi, \wt \rho)\,Q\,(\wt t, \wt \phi, \wt \rho)^T,
\ee
where the $Q$ operator is
\be
Q = \left(\begin{array}{ccc}
\partial_0^2+\partial_1^2-\mu_t^2 & 0 & -2\,i\,\kappa\,\sinh \rho\,\partial_0 \\
0 & -\partial_0^2-\partial_1^2+\mu_\phi^2 & 2\,i\,w\,\cosh\rho\,\partial_0 \\
2\,i\,\kappa\,\sinh \rho\,\partial_0 & -2\,i\,w\,\cosh\rho\,\partial_0 &  -\partial_0^2-\partial_1^2+\mu_\rho^2
\end{array}
\right) .
\ee
Since $t= \kappa \tau$
the  1-loop correction to string energy  is given by
\be \la{loi}
E_1 = {\G_1 \ov \k \T}  \ , \ \ \ \ \ \   \qquad \T \equiv \int d \tau \to \infty   \ . \ee
The effective action $\Gamma_1$ is computed in terms of the logarithm of a ratio of functional determinants. As was discussed in \cite{tt},
since the fluctuations Lagrangian does not depend on $\sigma$
we can reduce the two-dimensional functional determinants to one-dimensional ones (here $V(\sigma)$ is a generic operator) just by tracing over the temporal
dependence
\begin{equation}
\ln\det[-\partial_1^2 - \partial_0^2 + V(\sigma)]=
\T
\int {d{\omega}\ov 2\pi}  \ln\det [-\partial_1^2+\omega^2+ V(\sigma)] \ .
\end{equation}
Hence, the $1$-loop correction to the energy can be written as
\begin{equation}
\Gamma_1=-\ln Z_1=-\frac{\T}{4 \pi}\int_{-\infty}^{\infty}d \omega\ \ln \frac{
\det^{8}[-\partial_1^2+\omega^2+ \mu^2_F] \det[-\partial_1^2+\omega^2]}{\ \det^{2}[-\partial_1^2+\omega^2+\mu^2_{\beta}]\ \det^{4}[-\partial_1^2+\omega^2+\J^2]  \det[Q_{\omega}]} \ , \label{abs}
\end{equation}
where $Q_{\omega}=Q(\partial_0 \rightarrow i \omega)$. The determinant of the massless operator in the numerator
 comes from one of the two conformal gauge ghost contributions. The other ghost contribution cancels one of the massless mode from $S^5$.
The exact computation of these determinants is a challenging problem due to the complicated form of the solution $\rho=\rho(\sigma)$. In this paper we follow the prescription developed in \cite{tt} and compute the determinants in (\ref{abs}) perturbatively in small $\epsilon$ expansion.

\subsection{Correction at order $O(\epsilon^4)$}

We are interested in computing the $1$-loop correction to the energy in the limit $\sqrt \S \ll 1$ with $u=\frac{\J^2}{\S}$ fixed. To accomplish this we take $\epsilon$ to zero
while  keeping the parameter $x\equiv
\frac{\J}{\epsilon}$ fixed,  i.e. scaling $\J$ to zero
together with $\eps$  so that $\frac{\sqrt{u}}{2}={\J \ov \sqrt {2 \S}  } \approx x$
remains finite. Therefore, we will get the $1$-loop correction to the classical energy (\ref{zz1}). We can then  further expand in small $u$ and recover the
case of $\J \ll  \sqrt{\S} \ll 1$.
The $1$-loop correction in this particular case corresponds to the classical energy \rf{zz}.

Setting $\J= \epsilon x$, we proceed by expanding the fluctuation Lagrangian in small $\epsilon$. The parameters $\kappa, w$ and the string profile can be expanded as
\begin{eqnarray}
\kappa &=& \epsilon\,\sqrt{x^2+1}  -\frac{\epsilon ^3}{4 \sqrt{x^2+1}}+\oh{\epsilon ^5} \\
w &=& 1+\frac{1}{4} \left(2 x^2+1\right) \epsilon ^2- \frac{1}{64} \left[8 \left(x^4+x^2\right)+7\right] \epsilon ^4+\oh{\epsilon ^6}
\end{eqnarray}
\be
\rho(\sigma) = \epsilon\,\sin\sigma +\frac{\epsilon^3}{12}\,\sin\sigma\,(\sin^2\sigma-3) + \oh{\epsilon^5}.
\ee
The proportionality factors that appear in the coupled part of the fluctuation Lagrangian are expanded as
\begin{eqnarray}
\kappa\,\sinh\rho &=& \epsilon^2\,\sqrt{x^2+1} \sin\sigma-\epsilon^4\,\frac{\left(x^2 \cos 2\sigma +x^2+\cos 2\sigma+3\right) \sin\sigma}
{8 \sqrt{x^2+1}}+\oh{\epsilon^6}, \\
w\,\cosh\rho &=& 1+\frac{\epsilon^2}{4} \left(2 x^2+2 \sin ^2\sigma+1\right) +\frac{\epsilon^4}{64} \left(-8 x^4-8 x^2 \cos 2 \sigma +\cos 4 \sigma-8\right)+\oh{\epsilon^6}
\nonumber
\end{eqnarray}
The masses are
\begin{eqnarray}
\mu_t^2    &=& \epsilon^2 \cos 2 \sigma \epsilon ^2+\left(\frac{1}{4} \sin ^2 2 \sigma-\frac{1}{2} \cos 2 \sigma\right) \epsilon ^4+\cdots,\\
\mu_\phi^2 &=& -1+\left(\cos 2 \sigma+\frac{1}{2}\right) \epsilon ^2+\left(\frac{5}{32}-\cos ^4\sigma\right) \epsilon ^4+\cdots,\\
\mu_\rho^2 &=& -1+\left(\cos 2 \sigma-\frac{1}{2}\right) \epsilon ^2+\left(-\frac{1}{2} \cos 2 \sigma-\frac{1}{8} \cos 4 \sigma+\frac{9}{32}\right) \epsilon ^4+\cdots,\\
\mu_\beta^2&=& \left(x^2+\cos 2 \sigma+1\right) \epsilon ^2- \epsilon^4 \cos ^4\sigma +\cdots,\\
\mu_F^2    &=& \left(x^2+\cos ^2\sigma\right) \epsilon ^2-\frac{1}{2} \epsilon^4 \cos ^4\sigma +\cdots .
\end{eqnarray}
The expansion of the operator $Q_{\omega}$ has the form $Q_{\omega} = Q^{(0)}_{\omega} + \epsilon^2\,Q^{(2)}_{\omega} + \epsilon^4\,Q^{(4)}_{\omega} + \dots$ where
\be
Q^{(0)}_{\omega} = \left(
\begin{array}{lll}
 -n^2-\omega^2 & 0 & 0 \\
 0 & n^2+\omega^2-1 & -2 \omega \\
 0 & 2 \omega & n^2+\omega^2-1
\end{array}
\right) \ ,
\ee
\be
Q^{(2)}_{\omega} = \left(
\begin{array}{lll}
 -\cos (2 \sigma) & 0 & 2 \omega \sqrt{x^2+1} \sin \sigma \\
 0 & \cos (2 \sigma)+\frac{1}{2} & -\frac{1}{2} \omega \left(2 x^2+2 \sin ^2\sigma+1\right) \\
 -2 \omega \sqrt{x^2+1} \sin \sigma & \frac{1}{2} \omega \left(2 x^2+2 \sin ^2\sigma+1\right) & \cos 2 \sigma-\frac{1}{2}
\end{array}
\right) \ ,
\ee
\be
Q^{(4)}_{\omega} =
\left(
\begin{array}{lll}
 \frac{1}{4} \left(2 \cos (2 \sigma)-\sin ^2(2 \sigma)\right) & 0 & Q_{13} \\
 0 & \frac{5}{32}-\cos ^4\sigma & Q_{23} \\
 -Q_{13}& -Q_{23} &
   \frac{1}{32} (-16 \cos 2 \sigma-4 \cos 4 \sigma+9)
\end{array}
\right) \ .
\ee
where
\begin{equation}
Q_{13}=-\frac{\omega \left[x^2+\left(x^2+1\right) \cos 2 \sigma+3\right] \sin \sigma}{4 \sqrt{x^2+1}}, \quad
Q_{23}=-\frac{1}{32} \omega \left[\cos 4 \sigma-8 \left(x^4+x^2 \cos 2 \sigma +1\right)\right] \ .
\end{equation}
In the $\sigma$-independent leading operator $Q^{(0)}_{\omega}$ we have replaced $\partial_1 \rightarrow i n$ as suitable for periodic fields.
In order to compute perturbatively in $\epsilon$ the logarithm of the determinants appearing in (\ref{abs}) we use the expansion
\begin{eqnarray}
\ln  \det[1+\epsilon^2\,B+\epsilon^4\,C+\epsilon^6\,D] &=& \epsilon^2\,\mbox{Tr}\,B + \epsilon^4\,\left(\mbox{Tr}\,C-\frac{1}{2}\,\mbox{Tr}\,B^2\right) + \\
&&
+ \epsilon^6\,\left(\mbox{Tr}\,D-\mbox{Tr}\,(B\,C)+\frac{1}{3}\,\mbox{Tr}\,B^3\right) + \cdots \ ,  \label{trace}
\end{eqnarray}
where $B$ represents the product of the propagator at leading order in $\epsilon$ and the order $\epsilon^2$ expansion, i.e. for the coupled part $B= [Q_{\omega}^{(0)}]^{-1}\,Q^{(2)}_{\omega}$, and $C$ and $D$ are products of propagators and expansions in $\epsilon$ at orders $\epsilon^4$, and $\epsilon^6$ respectively.

At order $\epsilon^2$ the result was already obtained in \cite{tt}. It was showed that the $1$-loop correction to (\ref{zz1}) at order $\sqrt{S}$ vanishes. This is expected since at that order the string energy is the same as in flat space. Here we compute the $1$-loop correction to order $O(\epsilon^4)$.

To evaluate the traces in (\ref{trace}) we insert as in \cite{tt} a complete set of eigenstates of the ``unperturbed" operator  $Q_{\omega}^{(0)}$. In the case of the decoupled bosons and fermions these states are simply
\be
\langle\sigma | n\rangle = \frac{1}{\sqrt{2\,\pi}}\,e^{i\,n\,\sigma}.
\ee
In the case of the coupled $Q$ system, it is convenient to first rotate $(\wt t, \wt \phi, \wt \rho)$ in order to diagonalize $Q^{(0)}_{\omega}$. We do this by replacing
\be
Q^{(n)}_{\omega}\to R^{-1}\,Q^{(n)}_{\omega}\,R,
\ee
where
\be
R = \left(
\begin{array}{lll}
 1 & 0 & 0 \\
 0 & \frac{1}{2} & \frac{1}{2} \\
 0 & -\frac{i}{2} & \frac{i}{2}
\end{array}
\right)
\ee
In this basis, the eigenstates are defined as
\begin{eqnarray}
\langle\sigma | n\rangle_I &=& \frac{1}{\sqrt{2\,\pi}}\,e^{i\,n\,\sigma}\,(1,0,0)^T, \quad
\langle\sigma | n\rangle_{II} = \frac{1}{\sqrt{2\,\pi}}\,e^{i\,n\,\sigma}\,(0,1,0)^T, \\
\langle\sigma | n\rangle_{III} &=& \frac{1}{\sqrt{2\,\pi}}\,e^{i\,n\,\sigma}\,(0,0,1)^T.
\end{eqnarray}
After diagonalization the momentum-space propagator
corresponding to $Q_{\omega}^{(0)}$  is
\begin{eqnarray}
(Q_{\omega}^{(0)})^{-1} \rightarrow \left(
               \begin{array}{ccc}
                 -\frac{1}{n^2+\omega^2} & 0 & 0 \\
                 0 & \frac{1}{n^2+(\omega+i)^2} & 0 \\
                 0 & 0 & \frac{1}{n^2+(\omega-i)^2} \\
               \end{array}
             \right)    , \ \ \ \ \
 \la{my}
\end{eqnarray}
Since the propagator does not have standard massless form, we perform, as in \cite{tt}, $w$-shifts in denominators that contain $n^2+ (\omega \pm i)^2$ so that we put propagators in standard form, i.e. denominators $n^2+ \omega^2$. More precisely we write products of propagators as $\frac{1}{a\,b} = \frac{1}{a-b}\left(\frac{1}{b}-\frac{1}{a}\right)$, and we perform $\omega$-shifts in individual terms as needed.

Using (\ref{trace}) we compute the contribution to the logarithm of the partition functions from the various fields as they appear in (\ref{abs}).
The order $O(\epsilon^4)$ the result can be written as
\begin{eqnarray}
\Gamma_1 &=& -\frac{{\cal T}}{4\,\pi}\,X^{(4)}\,\epsilon^4 + O(\epsilon^6), \\
\label{eq:tmp1}
\kappa\,E_1 &=& \frac{\Gamma_1}{\cal T} =  -\frac{1}{4\,\pi}\,X^{(4)}\,\epsilon^4 + O(\epsilon^6) \ ,
\end{eqnarray}
where the quantity $X^{(4)}$ takes the form
\be
X^{(4)} = \int_{-\infty}^\infty d\omega\,\sum_{n\in\mathbb{Z}} X^{(4)}_n(\omega).
\ee

As in \cite{tt}, to isolate the modes constant in $\sigma$, we drop the $n=0$ contributions from the sums over $n$. Also, the prescription established in \cite{tt} is to perform the sum over modes first and then the integral over $\omega$ \footnote{If we perform the integral over $\omega$ first we get a different result. This is related to a regularization anomaly discussed in Appendix B in \cite{tt}.}. Implementing this prescription in a Mathematica code we find that remarkably the sums over $n$ can be done exactly and we obtain that the explicit value of the sum $X^{(4)}(\omega)$ is a polynomial in $x^2$ (some details of the computation are presented in Appendix A)
\be
X^{(4)}(\omega) = X^{(4,0)}(\omega) + X^{(4,2)}(\omega)\,x^2+X^{(4,4)}(\omega)\,x^4,
\ee
where
\begin{eqnarray}
X^{(4,0)}(\omega) &=& \frac{\pi ^2 \left(\omega ^2+1\right) {\rm csch}^2(\pi  \omega )}{2 \omega ^2}+\frac{\pi  \left(5 \omega ^2+4\right) \coth (\pi  \omega )}{8 \omega ^3
   \left(\omega ^2+1\right)}-\frac{1}{8 \left(\omega ^2+1\right)}+\nonumber\\
&& + \frac{7}{16 \left(\omega ^2+4\right)}-\frac{5}{8 \omega ^2}-\frac{1}{\left(\omega
   ^2+1\right)^2}-\frac{1}{\omega ^4},
\end{eqnarray}
\begin{equation}
X^{(4,2)}(\omega) = -\frac{1}{\left(\omega ^2+1\right)^2}, \quad \quad X^{(4,4)}(\omega) = \frac{1}{2 \omega ^2}-\frac{1}{2} \pi ^2 {\rm csch}^2(\pi  \omega ).
\end{equation}
Remarkably, the resulting integral over $\omega$ can also be computed exactly and we obtain
\be
X^{(4)} =  \int_{-\infty}^\infty d\omega\,\left[X^{(4,0)}(\omega) + X^{(4,2)}(\omega)\,x^2+X^{(4,4)}(\omega)\,x^4\right]= \pi \bigg(-\frac{41}{32}+\zeta_3-\frac{1}{2}\,x^2+x^4\bigg).
\ee

\subsection{Correction at order $O(\epsilon^6)$}

In this section we extend the above computation to order $O(\epsilon^6)$. This means that we need the expansions of parameters entering the fluctuations Lagrangian to higher order. Denoting by $\delta$ the corrections due to the next order contributions to various parameters needed,
we obtain the corrections to $\kappa$ and $w$
\begin{equation}
\delta\kappa = \frac{\left(11 x^2+9\right) \epsilon ^5}{64 \left(x^2+1\right)^{3/2}}, \quad \quad
\delta w = \frac{1}{256} \left(16 x^6+24 x^4+22 x^2+17\right) \epsilon ^6  \ ,
\end{equation}
the correction to the string solution
\be
\delta\rho = \frac{1}{320} \epsilon ^5 \left(4 \sin ^4\sigma-25 \sin ^2\sigma+45\right) \sin\sigma \ ,
\ee
the corrections to the mixing coefficients
\begin{equation}
\delta(\kappa\,\sinh\rho) = \frac{\epsilon ^6 \left(8 x^4+42 x^2+\left(x^2+1\right) \left(9 x^2+13\right) \cos 2 \sigma+\left(x^2+1\right)^2 \cos 4 \sigma+30\right) \sin\sigma}{128
   \left(x^2+1\right)^{3/2}}, \nonumber
\end{equation}
\begin{equation}
\delta(w\,\cosh\rho) = \frac{\epsilon ^6 \left(16 [\left(2 x^3+x\right)^2+4]+[32 (x^2+2) x^2+13] \cos 2 \sigma+8 \left(x^2-1\right) \cos 4 \sigma-
\cos 6\sigma\right)}{1024} \ ,
\nonumber
\end{equation}
and finally the corrections to the masses
\begin{eqnarray}
\delta \mu_t^2    &=& \frac{1}{256} \epsilon ^6 (85 \cos 2 \sigma+32 \cos 4 \sigma+3 \cos 6 \sigma-32), \\
\delta \mu_\phi^2 &=& \frac{1}{256} \epsilon ^6 (85 \cos 2 \sigma+32 \cos 4 \sigma+3 \cos 6 \sigma+36), \\
\delta \mu_\rho^2 &=& \frac{1}{256} \epsilon ^6 (85 \cos 2 \sigma+32 \cos 4 \sigma+3 \cos 6 \sigma-52), \\
\delta \mu_\beta^2&=& \frac{1}{64} \epsilon ^6 [26 \cos 2 \sigma+3 (\cos 4 \sigma+5)] \cos ^2\sigma , \\
\delta \mu_F^2    &=& \frac{1}{128} \epsilon ^6 [26 \cos 2 \sigma+3 (\cos 4 \sigma+5)] \cos ^2\sigma  \ .
\end{eqnarray}
As at order $O(\epsilon^4)$ we again use the expansion (\ref{trace}). At order $O(\epsilon^6)$ the most
difficult term is $\mbox{Tr}B^3$ which requires a double sum over virtual states. However, as usual, the
two infinite sums truncate to a finite sum since we need the products
\be
\langle n | B | n'\rangle\,\langle n' | B | n''\rangle\,\langle n'' | B | n\rangle\,
\ee
which are non vanishing only when the three conditions
\be
|n-n'| \leq  2, \quad |n'-n''| \leq 2, \quad |n''-n| \leq 2,
\ee
are all satisfied.

Let us write again the result us
\begin{eqnarray}
\Gamma_1 &=& -\frac{{\cal T}}{4\,\pi}\,+X^{(4)}\,\epsilon^4 +X^{(6)}\,\epsilon^6 + O(\epsilon^8), \\
\label{eq:tmp2}
\kappa\,E_1 &=& \frac\Gamma{\cal T} =  -\frac{1}{4\,\pi}\,X^{(4)}\,\epsilon^4-\frac{1}{4\,\pi}\,X^{(6)}\,\epsilon^6 +O(\epsilon^8). \label{abc}
\end{eqnarray}
The correction $X^{(6)}$ takes the form
\be
X^{(6)} = \int_{-\infty}^\infty d\omega\,\sum_{n\in\mathbb{Z}} X^{(6)}_n(\omega).
\ee

Again, it is crucial that we first compute the infinite sum
\be
X^{(6)}(\omega) = \sum_{n\in\mathbb{Z}} X^{(6)}_n(\omega),
\ee
and only as a final step integrate over $\omega$.

\medskip
The sum can be done exactly, and the explicit value of the sum $X^{(6)}(\omega)$ is a polynomial in $x^2$
\be
X^{(6)}(\omega) = X^{(6,0)}(\omega) + X^{(6,2)}(\omega)\,x^2+X^{(6,4)}(\omega)\,x^4+X^{(6,6)}(\omega)\,x^6,
\ee
where
\begin{eqnarray}
\lefteqn{
X^{(6,0)}(\omega) = -\frac{\pi ^3 \left(12 \omega ^4+5 \omega ^2+5\right) \coth ^3(\pi  \omega )}{24 \omega ^3}+ \coth(\pi\,\omega)\times } && \nonumber\\
&&  \frac{\pi  \left(192 \pi ^2 \omega ^{10}+464 \pi ^2 \omega
   ^8+\left(-411+432 \pi ^2\right) \omega ^6+\left(-746+240 \pi ^2\right) \omega ^4+\left(-499+80 \pi ^2\right) \omega ^2-120\right) }{384
   \omega ^5 \left(\omega ^2+1\right)^2} + \nonumber\\
&& -\frac{\pi ^2 \left(192 \omega ^6+473 \omega ^4+379 \omega ^2+120\right) \text{csch}^2(\pi  \omega )}{384 \omega ^4
   \left(\omega ^2+1\right)}-\frac{15}{16 \left(\omega ^2+1\right)}-\frac{269}{192 \left(\omega ^2+4\right)}+\frac{257}{192 \omega ^2}+\nonumber\\
&& + \frac{37}{8 \left(\omega
   ^2+1\right)^2}+\frac{517}{192 \left(\omega ^2+4\right)^2}-\frac{2}{\left(\omega ^2+1\right)^3}+\frac{299}{192 \omega ^4}+\frac{5}{6 \omega ^6},
\end{eqnarray}
\begin{eqnarray}
\lefteqn{
X^{(6,2)}(\omega) = -\frac{\pi ^3 \left(6 \omega ^4+\omega ^2+2\right) \coth ^3(\pi  \omega )}{6 \omega ^3}+\coth(\pi\,\omega)\times } \nonumber\\
&&  \frac{\pi  \left(192 \pi ^2 \omega ^{10}+416 \pi ^2 \omega
   ^8+\left(-147+320 \pi ^2\right) \omega ^6+4 \left(-69+40 \pi ^2\right) \omega ^4+\left(-273+64 \pi ^2\right) \omega ^2-96\right)}{192
   \omega ^5 \left(\omega ^2+1\right)^2} + \nonumber\\
&& -\frac{\pi ^2 \left(35 \omega ^4+59 \omega ^2+32\right) \text{csch}^2(\pi  \omega )}{64 \omega ^4 \left(\omega
   ^2+1\right)}-\frac{41}{16 \left(\omega ^2+1\right)}-\frac{31}{32 \left(\omega ^2+4\right)}+\frac{39}{32 \omega ^2}+\frac{65}{8 \left(\omega
   ^2+1\right)^2}+\nonumber\\
&& + \frac{91}{32 \left(\omega ^2+4\right)^2}-\frac{3}{\left(\omega ^2+1\right)^3}+\frac{97}{96 \omega ^4}+\frac{4}{3 \omega ^6},
\end{eqnarray}
\begin{eqnarray}
X^{(6,4)}(\omega) &=& \frac{\pi ^3 \left(-12 \omega ^4+7 \omega ^2+5\right) \coth ^3(\pi  \omega )}{24 \omega ^3}+\frac{\pi  \left(24 \pi ^2 \omega ^6-14 \pi ^2 \omega ^4-\left(3+10
   \pi ^2\right) \omega ^2+15\right) \coth (\pi  \omega )}{48 \omega ^5}+\nonumber\\
&& +\frac{\pi ^2 \left(4 \omega ^4-\omega ^2+5\right) \text{csch}^2(\pi  \omega )}{16
   \omega ^4}-\frac{3}{2 \left(\omega ^2+1\right)}+\frac{1}{4 \omega ^2}+\frac{4}{\left(\omega ^2+1\right)^2} +\nonumber\\
&& -\frac{1}{\left(\omega ^2+1\right)^3}-\frac{1}{6
   \omega ^4}-\frac{5}{6 \omega ^6},
\end{eqnarray}
\begin{eqnarray}
X^{(6,6)}(\omega) &=& \frac{\pi ^3 \left(3 \omega ^2+1\right) \coth ^3(\pi  \omega )}{12 \omega ^3}+\frac{\pi  \left(-6 \pi ^2 \omega ^4+\left(3-2 \pi ^2\right) \omega ^2+3\right)
   \coth (\pi  \omega )}{24 \omega ^5}+\nonumber\\
&& + \frac{\pi ^2 \left(2 \omega ^4+\omega ^2+1\right) \text{csch}^2(\pi  \omega )}{8 \omega ^4}-\frac{1}{4 \omega
   ^2}-\frac{1}{2 \omega ^4}-\frac{1}{3 \omega ^6}.
\end{eqnarray}

We present details of the computation of the $\omega$-integrals in Appendix B. It is remarkable that again the integrals can be done exactly. The result is finite as expected from the expansion about any string solution \cite{dgt}. Collecting the result from all terms we obtain
\begin{eqnarray}
X^{(6)} &= & \int_{-\infty}^\infty d\omega\,\left[X^{(6,0)}(\omega) + X^{(6,2)}(\omega)\,x^2+X^{(6,4)}(\omega)\,x^4+X^{(6,6)}(\omega)\,x^6\right]\nonumber\\
&=& \pi \bigg[\frac{1343}{1024}-\frac{179}{192}\,\zeta_3-\frac{5}{8}\,\zeta_5 +
\left(
\frac{499}{512}-\frac{49}{96}\,\zeta_3-\zeta_5
\right)\,x^2 + \nonumber\\
&& + \left(
\frac{1}{8}-\frac{17}{24}\,\zeta_3+\frac{5}{8}\,\zeta_5
\right)\,x^4 +
\left(
-\frac{1}{2}-\frac{1}{4}\,\zeta_3+\frac{1}{4}\,\zeta_5
\right)\,x^6\bigg].  \label{kri}
\end{eqnarray}

We need now to express the result in terms of fixed $u$ and expand in small $\S$. The expression of $\kappa$ in this limit is
\begin{equation}
\kappa=\sqrt{u+2}\sqrt{\S}-\frac{2 u+5}{4 \sqrt{u+2}}\S^{\frac{3}{2}}+\frac{123+2 u(83+ 38 u+6 u^2)}{32 (u+2)^{\frac{3}{2}}}\S^{\frac{5}{2}}+O(\S^{\frac{7}{2}})  \ . \label{kwo}
\end{equation}

Replacing back $x=\frac{\J}{\epsilon}$ in the $1$-loop correction result (\ref{abc},\ref{kri}), then replacing everything in terms of the fixed $u$ and using the expression for $\epsilon$ in (\ref{zz2}) and $\kappa$ in (\ref{kwo}) we obtain the the $1$-loop correction to the energy (expanding in small $\S$)
\begin{equation}
E_1= \frac{41 - 8 u(u-1)-32 \zeta(3)}{32 \sqrt{u+2}}\S^{\frac{3}{2}}+\frac{A_0+A_1 u +A_2 u^2+A_3 u^3+A_4 u^4}{1536 (u+2)^{\frac{3}{2}}}\S^{\frac{5}{2}}+O(\S^{\frac{7}{2}}) \ .
\end{equation}
where
\begin{eqnarray}
A_0&=& -7566 + 5344 \zeta(3)+3840 \zeta(5), \quad A_1=-10671+7504 \zeta(3)+4992 \zeta(5),\nonumber\\
A_2&=&-4425 + 3408 \zeta(3)+576 \zeta(5), \quad
A_3= -96 +736 \zeta(3)- 672 \zeta(5), \nonumber\\
 A_4&=& 192 +96 \zeta(3)-96 \zeta(5) \ .
\end{eqnarray}
This result is the $1$-loop correction to the classical energy (\ref{zz1}).
In the small $u$ limit the correction becomes
\begin{eqnarray}
E_1&=& \S^{\frac{3}{2}}\bigg[\frac{41}{32 \sqrt{2}}-\frac{\zeta(3)}{\sqrt{2}}+\frac{\J^2}{\S}\bigg(\frac{\zeta(3)}{4 \sqrt{2}}-\frac{9}{128 \sqrt{2}}\bigg)+O(\frac{\J^4}{\S^2})\bigg]
+ \S^{\frac{5}{2}}\bigg[-\frac{1261}{512 \sqrt{2}}+ \frac{167 \zeta(3)}{96 \sqrt{2}}\nonumber\\
&+& \frac{5 \zeta(5)}{4 \sqrt{2}}
+\frac{\J^2}{\S}\bigg(-\frac{3331}{2048 \sqrt{2}}+\frac{437 \zeta(3)}{384 \sqrt{2}}+\frac{11 \zeta(5)}{16 \sqrt{2}}\bigg)+O(\frac{\J^4}{\S^2})\bigg]+O(\S^{\frac{7}{2}}) \ .
\end{eqnarray}
This represents the $1$-loop correction to (\ref{zz}). It is obvious that the structure of the classical expression (\ref{zz}) in this limit is preserved at the quantum level.

Let us consider the limit of the $1$-loop correction in large $u$ limit. Writing this correction together with the classical expression (\ref{zz3}) we obtain
\begin{eqnarray}
E&=&E_0+E_1=J \bigg(1+ \sqrt{\lambda}\frac{S}{J^2}\big(1+ 0) - \lambda \frac{S^2}{2 J^4}(1+0) + O(\frac{S^3}{J^6})\bigg)\nonumber\\
&-&\frac{1}{4}\frac{J^3}{\lambda^{\frac{3}{2}}}+  \frac{J S}{2 \sqrt{\lambda}}\bigg(1+ \frac{1}{\sqrt{\lambda}}+ O(\frac{1}{\lambda})\bigg)+
\frac{S^2}{4 J}\bigg[1+ \big(\frac{21}{8}-4 \zeta(3)\big) \frac{1}{\sqrt{\lambda}}+ O(\frac{1}{\lambda})\bigg]+O(\frac{S^4}{J^5})\nonumber\\
&+&\frac{2+ \zeta(3)-\zeta(5)}{16}\frac{J^5}{\lambda^{\frac{5}{2}}}- \frac{S J^3}{8 \lambda^{\frac{3}{2}}}\bigg[1+\big(\frac{7}{2}-\frac{7 \zeta(3)}{3 } +2 \zeta(5)\big) \frac{1}{\sqrt{\lambda}}+O(\frac{1}{\lambda})\bigg]\nonumber\\
&-&\frac{S^2 J}{4 \lambda} \bigg[1+ \big(\frac{899}{128}-5 \zeta(3)-\frac{39}{8} \zeta(5)\big)+O(\frac{1}{\lambda})\bigg]+O(\frac{S^3}{J}) \label{zz4}
\end{eqnarray}
where the first line is just the classical energy coming from the leading flat-space type energy. The zeros indicate that that part does not receive quantum corrections. We observe the fact that in contrast to the small $u$ limit, now the structure of the $1$-loop corrections is different than that at the classical level by terms like $J^3 \lambda^{-\frac{3}{2}}, J^5 \lambda^{-\frac{5}{2}},...$, which start appearing at $1$-loop. As was already mentioned, this is not a BMN type expansion since here a different limit is considered.

\bigskip

\section*{Acknowledgments }

We thank  A. Tseytlin and G.~P.~Korchemsky for
useful  discussions and helpful comments on the draft.
A.T. was supported in part by NSF under grant PHY-0653357.

\appendix
\subsection*{Appendix A: Separate contributions to $X^{(4,a)}$ and finiteness
}
\label{app:partial}

\refstepcounter{section}
\def\theequation{A.\arabic{equation}}
\setcounter{equation}{0}

Each term $X^{(4,a)}$ can be written as the sum of four contributions
\be
X^{(4,a)} = X^{(4,a)}_\psi + X^{(4,a)}_\beta + X^{(4,a)}_\Phi + X^{(4,a)}_Q,
\ee
which come from fermions, decoupled bosons $\beta_i$ and $\Phi\equiv (\varphi, \chi_s)$, and coupled three bosons $Q$. All the contributions take into account
the sign from Fermi-Bose statistics as well as the multiplicity of the various fields. The separate contributions are

\subsection{$X^{(4,0)}$}

\begin{eqnarray}
X^{(4,0)}_\psi &=& -\frac{\pi ^2 {\rm csch}^2(\pi  \omega )}{2 \omega ^2}-\frac{\pi  \left(6 \omega ^4+9 \omega ^2+2\right) \coth (\pi  \omega )}{4 \omega ^3 \left(\omega
   ^2+1\right)}-\frac{1}{4 \left(\omega ^2+4\right)}+\frac{7}{4 \omega ^2}+\frac{1}{\omega ^4}, \nonumber \\
X^{(4,0)}_\beta&=& \frac{\pi ^2 {\rm csch}^2(\pi  \omega )}{2 \omega ^2}+\frac{\pi  \left(3 \omega ^4+6 \omega ^2+2\right) \coth (\pi  \omega )}{4 \omega ^3 \left(\omega
   ^2+1\right)}+\frac{1}{4 \left(\omega ^2+4\right)}-\frac{1}{\omega ^2}-\frac{1}{\omega ^4}, \nonumber \\
X^{(4,0)}_\Phi &=& 0, \nonumber\\
X^{(4,0)}_Q    &=& \frac{\pi ^2 \left(\omega ^2+1\right) {\rm csch}^2(\pi  \omega )}{2 \omega ^2}+\frac{\pi  \left(6 \omega ^4+11 \omega ^2+4\right) \coth (\pi  \omega )}{8
   \omega ^3 \left(\omega ^2+1\right)} + \\
&& -\frac{1}{8 \left(\omega ^2+1\right)}+\frac{7}{16 \left(\omega ^2+4\right)}-\frac{11}{8 \omega ^2}-\frac{1}{\left(\omega
   ^2+1\right)^2}-\frac{1}{\omega ^4}. \nonumber
\end{eqnarray}
The IR $\omega=0$ singularities cancel in all separate terms as a consequence of the zero mode projection. The UV large $\omega$ behaviour is instead
\be
X^{(4,0)}_\psi \sim -\frac{3 \pi }{2 \omega } + \dots, \quad
X^{(4,0)}_\beta\sim \frac{3 \pi }{4 \omega }+\dots, \quad
X^{(4,0)}_\Phi \sim 0, \quad
X^{(4,0)}_Q    \sim \frac{3 \pi }{4 \omega } + \dots,
\ee
with UV finiteness $X^{(4,0)} \sim {\cal O}(\omega^{-2})$.

\subsection{$X^{(4,2)}$}

\begin{eqnarray}
X^{(4,2)}_\psi &=& -\frac{2 \pi ^2 {\rm csch}^2(\pi  \omega )}{\omega ^2}-\frac{2 \pi  \coth (\pi  \omega )}{\omega ^3}+\frac{4}{\omega ^4}, \\
X^{(4,2)}_\beta&=& \frac{\pi ^2 {\rm csch}^2(\pi  \omega )}{\omega ^2}+\frac{\pi  \coth (\pi  \omega )}{\omega ^3}-\frac{2}{\omega ^4}, \nonumber\\
X^{(4,2)}_\Phi &=& 0, \nonumber\\
X^{(4,2)}_Q    &=& \frac{\pi ^2 {\rm csch}^2(\pi  \omega )}{\omega ^2}+\frac{\pi  \coth (\pi  \omega )}{\omega ^3}-\frac{1}{\left(\omega ^2+1\right)^2}-\frac{2}{\omega ^4}.
\end{eqnarray}
The IR $\omega=0$ singularities cancel in all separate terms as a consequence of the zero mode projection. The UV large $\omega$ behaviour is
at least ${\cal O}(\omega^{-2})$ for all terms and UV finiteness is assured.

\subsection{$X^{(4,4)}$}

\begin{eqnarray}
X^{(4,4)}_\psi &=& -\frac{2 \pi ^2 {\rm csch}^2(\pi  \omega )}{\omega ^2}-\frac{2 \pi  \coth (\pi  \omega )}{\omega ^3}+\frac{4}{\omega ^4}, \\
X^{(4,4)}_\beta&=& \frac{\pi ^2 {\rm csch}^2(\pi  \omega )}{2 \omega ^2}+\frac{\pi  \coth (\pi  \omega )}{2 \omega ^3}-\frac{1}{\omega ^4},\nonumber\\
X^{(4,4)}_\Phi &=& \frac{\pi ^2 {\rm csch}^2(\pi  \omega )}{\omega ^2}+\frac{\pi  \coth (\pi  \omega )}{\omega ^3}-\frac{2}{\omega ^4}, \nonumber\\
X^{(4,4)}_Q    &=& -\frac{\pi ^2 \left(\omega ^2-1\right) {\rm csch}^2(\pi  \omega )}{2 \omega ^2}+\frac{\pi  \coth (\pi  \omega )}{2 \omega ^3}+\frac{1}{2 \omega
   ^2}-\frac{1}{\omega ^4}.\nonumber
\end{eqnarray}
The IR $\omega=0$ singularities cancel in all separate terms as a consequence of the zero mode projection. The UV large $\omega$ behaviour is
at least ${\cal O}(\omega^{-2})$ for all terms and UV finiteness is assured.

\newpage
\section{Analytical evaluation of the $\omega$ integrals}

\refstepcounter{section}
\def\theequation{B.\arabic{equation}}
\setcounter{equation}{0}

We want to compute general integrals of the form
\be
\int_\mathbb{R} d\omega \, f(\omega),
\ee
where $f(\omega)$ is in the class of the various $X^{(a,b)}$ contributions, i.e. rational functions of $\omega$ times possible
$\coth(\pi\omega)$ or ${\rm cshc}(\pi\omega)$ factors. The rational functions have possible poles at $\omega = \pm i, \pm 2\,i$.
The general recipe that we now discuss works for all the cases of interest.

We split $f(\omega)$ as
\be
f(\omega) = f_0(\omega) + f_1(\omega),
\ee
where $f_0(\omega)$ has a vanishing integral over the upper half circumference
\be
\Gamma_+ = \{R\,e^{i\,\vartheta},\ 0 \leq \vartheta \leq \pi\},
\ee
in the $R\to +\infty$ limit, and $f_1(\omega)$ is uniquely identified by looking at the large $\omega$ behaviour of $f(\omega)$ as shown in the
examples. Both $f_{0,1}$ are taken to be regular at $\omega=0$ by addition and subtraction of a suitable pole at $\omega=0$.

After this splitting, by contour deformation, we immediately obtain
\be
\int_\mathbb{R} d\omega \, f(\omega) = 2\,\pi\,i\,\sum_{n=1}^\infty \mbox{Res}_{\omega = i\,n} f_0(\omega) - \int_{\Gamma_+} d\omega\, f_1(\omega).
\ee
In all cases, we can write $f_1(\omega) = F'_1(\omega)$ for a simple function $F_1(\omega)$ which can be given in closed form. Therefore,
\be
\int_\mathbb{R} d\omega \, f(\omega) = 2\,\pi\,i\,\sum_{n=1}^\infty \mbox{Res}_{\omega = i\,n} f_0(\omega) - F_1(-\infty) + F_1(+\infty).
\ee
We now give the detailed evaluation of $X^{(4,a)}$ and $X^{(6,a)}$.

\subsection{$X^{(4,0)}$}

We take
\be
f_1(\omega) = \frac{1}{2} \pi ^2 {\rm csch}^2(\pi  \omega )-\frac{1}{2 \omega ^2},
\ee
and therefore
\be
 - \int_{\Gamma_+} d\omega\, f_1(\omega) = -\pi.
\ee
The sum over residues is
\be
\int_\mathbb{R} d\omega \, f_0(\omega) = \frac{9 \pi }{16}+\frac{29 \pi }{96}+\sum_{n\ge 3} \frac{\left(3 n^2-4\right) \pi }{4 n^3 \left(n^2-1\right)} =
-\frac{9 \pi }{32}+\pi  \zeta_3.
\ee
Summing the $f_1$ contribution we get
\be
\int_\mathbb{R} d\omega \, f(\omega) = \pi  \left(-\frac{41}{32}+\zeta_3\right).
\ee

\subsection{$X^{(4,2)}$ and $X^{(4,4)}$}

The $\omega$-integration of these terms is elementary. We get
\be
X^{(4,2)} = \int_\mathbb{R} d\omega\, \left[-\frac{1}{\left(\omega ^2+1\right)^2}\right] = -\frac{\pi}{2},
\ee
and
\be
X^{(4,4)} = \int_\mathbb{R} d\omega\, \left[ \frac{1}{2 \omega ^2}-\frac{1}{2} \pi ^2 {\rm csch}^2(\pi  \omega )\right] =
\int_\mathbb{R} d\omega\, \frac{d}{d\omega}\left[\frac{1}{2} \pi  \coth (\pi  w)-\frac{1}{2 w}\right] = \pi.
\ee

\subsection{$X^{(6,0)}$}

We take
\be
f_1(\omega) = \frac{1}{\omega ^2}-\frac{1}{2} \pi ^2 (\pi  \omega  \coth (\pi  \omega )+1) {\rm csch}^2(\pi  \omega ),
\ee
and therefore
\be
 - \int_{\Gamma_+} d\omega\, f_1(\omega) = \frac{3 \pi }{2}.
\ee
The sum over residues is
\be
\int_\mathbb{R} d\omega \, f_0(\omega) = -\frac{65 \pi }{48}-\frac{1765 \pi }{3072}+\sum_{n\ge 3} \frac{\left(-71 n^4+59 n^2+120\right) \pi }{192 n^5 \left(n^2-1\right)}
 = -\frac{193 \pi }{1024}-\frac{179}{192} \pi  \zeta_3-\frac{5}{8} \pi  \zeta_5.  \nonumber
\ee
Summing the $f_1$ contribution we get
\be
\int_\mathbb{R} d\omega \, f(\omega) = \pi\,\left(\frac{1343}{1024}-\frac{179}{192}\,  \zeta_3-\frac{5}{8} \,  \zeta_5\right).
\ee

\subsection{$X^{(6,2)}$}

We take
\be
f_1(\omega) = \frac{1}{\omega ^2}-\pi ^3 \omega  \coth (\pi  \omega ) {\rm csch}^2(\pi  \omega ),
\ee
and therefore
\be
 - \int_{\Gamma_+} d\omega\, f_1(\omega) = \pi.
\ee
The sum over residues is
\begin{eqnarray}
\int_\mathbb{R} d\omega \, f_0(\omega) = -\frac{245 \pi }{192}-\frac{569 \pi }{1536} -\sum_{n\ge 3} \frac{\left(31 n^4+47 n^2-96\right) \pi }{96 n^5 \left(n^2-1\right)}
= -\frac{13 \pi }{512}-\frac{49}{96} \pi  \zeta_3-\pi  \zeta_5.
\end{eqnarray}
Summing the $f_1$ contribution we get
\be
\int_\mathbb{R} d\omega \, f(\omega) = \pi  \left(\frac{499}{512}-\frac{49 \zeta_3}{96}-\zeta_5\right).
\ee

\subsection{$X^{(6,4)}$}

We take
\be
f_1(\omega) = \frac{1}{4 \omega^2}-\frac{1}{4} \pi ^2 {\rm csch}^3(\pi  \omega) (2 \pi  \omega \cosh (\pi  \omega)-\sinh (\pi  \omega)),
\ee
and therefore
\be
 - \int_{\Gamma_+} d\omega\, f_1(\omega) = 0.
\ee
The sum over residues gives the final result
\be
\int_\mathbb{R} d\omega \, f_0(\omega) = \frac{\pi }{24}+\sum_{n\ge 2} \frac{\left(15-17 n^2\right) \pi }{24 n^5} =
\pi  \left(\frac{1}{8}-\frac{17 \zeta_3}{24}+\frac{5 \zeta_5}{8}\right).
\ee

\subsection{$X^{(6,6)}$}

We take
\be
f_1(\omega) = \frac{1}{4} \pi ^2 {\rm csch}^2(\pi  \omega )-\frac{1}{4 \omega ^2},
\ee
and therefore
\be
 - \int_{\Gamma_+} d\omega\, f_1(\omega) = -\frac{\pi}{2}.
\ee
The sum over residues is
\be
\int_\mathbb{R} d\omega \, f_0(\omega) = -\sum_{n\ge 1} \frac{\left(n^2-1\right) \pi }{4 n^5} = -\frac{1}{4} \pi  \zeta_3+\frac{1}{4} \pi  \zeta_5.
\ee
Summing the $f_1$ contribution we get
\be
\int_\mathbb{R} d\omega \, f(\omega) = \pi  \left(-\frac{1}{2}-\frac{\zeta_3}{4}+\frac{\zeta_5}{4}\right).
\ee

\end{document}